\documentclass[11pt,preprint]{aastex}

\usepackage{amsmath}
\usepackage{mathptm}
\slugcomment{\it Accepted by ApJ, 03/01/11.}

\newcommand\HII{H\,{\sc ii}}

\newcommand\kms{km~s$^{-1}$}
\newcommand\cmt{cm$^{-2}$}
\newcommand\cc{cm$^{-3}$}

\newcommand\mum{$\mu$m}

\newcommand\mjb{mJy~beam$^{-1}$}
\newcommand\jb{Jy~beam$^{-1}$}

\begin{document}

\title{VLA Observations of the Infrared Dark Cloud G19.30+0.07}

\author{K. E. Devine\altaffilmark{1,2,3}, C. J. Chandler\altaffilmark{2},
C. Brogan\altaffilmark{4}}
\author{E. Churchwell\altaffilmark{1}, R. Indebetouw\altaffilmark{4,5},
Y. Shirley\altaffilmark{6}, K. J. Borg\altaffilmark{2,7}}
\altaffiltext{1}{Department of Astronomy, University of Wisconsin,
475 N. Charter Street, Madison, WI 53703}
\altaffiltext{2}{National Radio Astronomy Observatory, PO Box O, Socorro
NM 87801}
\altaffiltext{3}{Current address: Department of Physics, The
College of Idaho, 2112 Cleveland Blvd., Caldwell, ID 83605;
kdevine@collegeofidaho.edu}
\altaffiltext{4}{National Radio Astronomy Observatory, 520 Edgemont Road,
Charlottesville, VA 22903}
\altaffiltext{5}{Department of Astronomy, University of Virginia, PO
Box 400325, Charlottesville, VA 22903}
\altaffiltext{6}{Steward Observatory, University of Arizona, 933 N. Cherry
Ave., Tucson, AZ 85721}
\altaffiltext{7}{Current address: 6047 Valhalla Ave., Pensacola, FL 32507}

\begin{abstract}

We present Very Large Array observations of ammonia (NH$_3$) (1,1),
(2,2), and CCS ($2_1-1_0$) emission toward the Infrared Dark Cloud (IRDC)
G19.30+0.07 at $\sim 22$~GHz.  The NH$_3$ emission closely follows the
8~\mum\ extinction.  The NH$_3$ (1,1) and (2,2) lines provide diagnostics
of the temperature and density structure within the IRDC, with typical
rotation temperatures of $\sim 10$ to 20~K and NH$_3$ column densities
of $\sim 10^{15}$~cm$^{-2}$.  The estimated total mass of G19.30+0.07
is $\sim$1130~M$_{\sun}$.  The cloud comprises four compact NH$_3$
clumps of mass $\sim 30$ to 160~M$_{\sun}$.  Two coincide with 24~\mum\
emission, indicating heating by protostars, and show evidence of outflow
in the NH$_3$ emission.  We report a water maser associated with a
third clump; the fourth clump is apparently starless.  A non-detection
of 8.4~GHz emission suggests that the IRDC contains no bright \HII\
regions, and places a limit on the spectral type of an embedded ZAMS
star to early-B or later.  From the NH$_3$ emission we find G19.30+0.07
is composed of three distinct velocity components, or ``subclouds."
One velocity component contains the two 24~\mum\ sources and the starless
clump, another contains the clump with the water maser, while the third
velocity component is diffuse, with no significant high-density peaks.
The spatial distribution of NH$_3$ and CCS emission from G19.30+0.07 is
highly anti-correlated, with the NH$_3$ predominantly in the high-density
clumps, and the CCS tracing lower-density envelopes around those clumps.
This spatial distribution is consistent with theories of evolution for
chemically young low-mass cores, in which CCS has not yet been processed
to other species and/or depleted in high-density regions.

\end{abstract}


\section{Introduction}

Infrared dark clouds (IRDCs) are dense, cold molecular clouds seen
in silhouette against 8~\mum\ Galactic plane PAH emission.  Dust is
relatively transparent to 8~\mum\ emission, so only the densest parts
of molecular clouds (A$_V$ $\gtrsim 50$) are opaque at this wavelength
\citep{indebetouw}.  Since massive stars and their associated clusters
form in massive and dense concentrations of dust and gas, it follows that
the initial conditions of massive star formation are probably found in
the highest density regions of molecular clouds.  The GLIMPSE team imaged
about two-thirds of the inner Galactic plane $|l|=0^{\circ}-65^{\circ}$
using the {\em Spitzer Space Telescope (SST)} Infrared Array Camera
(IRAC) at 3.6, 4.5, 5.8, and 8~\mum\ with better than $2\arcsec$
resolution \citep{benjamin}.  The GLIMPSE images contain thousands of
IRDCs \citep{peretto} with structure on size scales $\ga 0.5\arcmin$
to as small as the IRAC resolution (see Figure~\ref{3col19}).  Because
they are prominent in silhouette, IRDCs are generally expected to lie
at the near kinematic distance.  Preliminary studies show that IRDCs
are cold ($\la 20$~K) and dense ($n_{\rm H_2} \ga 10^{5}$~\cc), and
many have masses $\ga 10^3$~M$_{\sun}$ \citep{carey98, egan98, pillai,
redman, rathborne06, rathborne07, wang08}.  \citet{pillai2} detected high
deuteration in IRDC clumps, evidence of very low temperatures. However,
the details of the kinematics, chemistry, and star formation within IRDCs
are not well understood.  High resolution observations of IRDCs are needed
to provide information on scales small enough to differentiate between
cloud properties in quiescent and star forming regions, a distinction
crucial to refining models of the earliest stages of massive star
formation \citep{wang08, rathborne06}.

Here we report results of a detailed study of an IRDC at radio
wavelengths using the Very Large Array (VLA) of the National
Radio Astronomy Observatory\footnote{The National Radio Astronomy
Observatory is a facility of the National Science Foundation
operated under cooperative agreement by Associated Universities,
Inc.}.  IRDC G19.30+0.07 ($\alpha$(J2000)=18$^h$25$^m$56$^s$,
$\delta$(J2000)=$-$12$^{\circ}$04\arcmin30\arcsec, V$_{LSR} \sim 26$~\kms)
is a filamentary IRDC with a length of $\sim 2$\arcmin\ and a width
of $\sim 30$\arcsec.  Using a flat Galactic rotation curve with R$_0 =
8.5$~kpc and V$_0 = 220$~\kms, the kinematic distance to G19.30+0.07 is
$\sim 2.4$~kpc.  We have determined the physical and chemical properties
of G19.30+0.07 using ammonia (NH$_3$) (1,1), (2,2), and dicarbon sulfide
(CCS) ($2_1-1_0$) transitions (upper level energies ($E_u/k$) of 23.8~K,
65.0~K, and 1.6~K respectively) with an angular resolution roughly a
factor of seven higher than previously reported for this IRDC, 6\arcsec\
compared to $\sim 40$\arcsec\ \citep{pillai, sakai08}.  To relate the
gas properties to star formation, continuum observations at 8.4 GHz
have been carried out to search for ionized gas associated with \HII\
regions in the IRDC.

NH$_{3}$ and CCS are important molecules for investigating IRDCs:
both are typically considered to be high-density tracers ($n_{\rm H_2}
\ga 10^{4-6}$~\cc), and the relative strengths of NH$_3$ inversion
lines can be used to estimate the temperature and column density of the
gas \citep{ho83}.  Furthermore, [CCS]/[NH$_3$] relative abundances may
provide information about chemical evolution \citep{suzuki, vandish}.
Chemical models have mainly been developed from observations of low mass
star forming regions; this study examines whether such models might also
apply in IRDCs near young star forming clumps.

The main goal of this paper is to examine the physical and chemical
properties of the IRDC G19.30+0.07 and explore the relationship between
these properties and signs of star formation.  The high spatial and
spectral resolution of this study permits us to distinguish between
multiple velocity components in the IRDC, map the temperature and
density distributions in the IRDC at 6\arcsec\ resolution, and examine
the relative distribution of two gas species in the IRDC.  This is one of
the first IRDC studies to explore the role of multiple cloud components,
distinguished by small offsets in velocity, within an IRDC.  Furthermore,
evidence for star formation and outflows in G19.30+0.07 enables us to
examine the relationship between IRDC properties and star formation.

This paper is structured as follows: in \S2, we present details of
our VLA observations and data reduction.  In \S3, we present results
of the NH$_3$, CCS, and continuum observations.  In \S4 we examine
the gas kinematics and present the derived temperature, density, and
mass distributions, as well as relative abundances of NH$_3$ and CCS
in the IRDC.  In \S5 we discuss the relationship between physical and
chemical properties, kinematics, and star formation within the IRDC.
Conclusions are given in \S6.

\section{Observations and Calibration}

Table~\ref{obssum} summarizes the parameters of the VLA observations.
NH$_3$ (1,1) and (2,2) and CCS ($2_1-1_0$) emission at $\sim 22$~GHz
were observed with the VLA in D configuration.  The 3.125~MHz bandwidth
contained the main and innermost satellite hyperfine lines of the NH$_3$
(1,1) and (2,2) transitions.  Two pointings were used to fully cover
the highest opacity regions, as determined from the 8~\mum\ images.
All data were calibrated and imaged using AIPS\@.  The CCS observations
were conducted during the EVLA upgrade, and eight of the twenty-seven
antennas had been converted to EVLA antennas.  Data were obtained using
both the VLA and EVLA antennas.  Doppler tracking could not be used
during the upgrade period, so the AIPS task CVEL was applied during
calibration to correct for motion of the Earth relative to the local
standard of rest during the observations.  In the 22~GHz datasets,
the uncertainty of the absolute flux densities is estimated to be 10\%.
We estimate the absolute positional accuracy to be better than 1\arcsec.

Images were made of the two pointings at each frequency using natural
weighting of the \emph{u-v} data and CLEAN deconvolution.  The AIPS
task FLATN was used to mosaic the two pointings and apply a primary
beam correction.  The NH$_3$ (1,1) and (2,2) mosaics were cut off at
the 50\% power level of the primary beam.  The synthesized beam in
the full-resolution NH$_3$ images is $\sim 4$\arcsec; however, the
NH$_3$ data were later smoothed to 6\arcsec\ resolution to increase
the signal-to-noise ratio prior to fitting the emission lines with
Gaussian profiles as a function of velocity (see \S4).  The multiplication
factors needed to convert from Jy~beam$^{-1}$ to beam-averaged brightness
temperature (K) for the various images are listed in Table~\ref{obssum}.

Because the CCS emission is weak and extended relative to that of NH$_3$,
a 30~k$\lambda$ Gaussian \emph{u-v} taper was applied while imaging
the data, resulting in a resolution of $7\farcs7 \times 5\farcs7$.  The
\emph{u-v} taper was chosen to maintain the highest resolution possible
while still providing a significant ($\ge 3\sigma$) CCS detection.
Additionally, a more conservative primary beam cutoff at the 65\%
power point was used while mosaicing the CCS so that noise peaks at the
edges of the primary beam were not amplified into apparently significant
detections at the low contour levels.  The 65\% power point cutoff was
determined by reducing the primary beam cutoff until spurious $3\sigma$
detections no longer appeared at the image edges.

Continuum emission at 8.4~GHz was observed with the VLA in A
configuration.  These data were also calibrated and imaged with the
AIPS reduction package, using the model for the flux calibrator supplied
with the package.  The uncertainty of the absolute flux density scale is
estimated to be 10\%.  Robust \emph{u-v} weighting and CLEAN deconvolution
were used during imaging.  The resolution of the resulting image is
$\sim 0\farcs3$.

We have also re-reduced and analyzed the G19.30+0.07 VLA H$_2$O maser
data published by \citet{wang} (project AW659; the full details of
these observations are described by Wang et al.).  The data for the
two pointings toward G19.30+0.07 were reduced in the usual manner
using AIPS\@.  After self-calibration, the RMS noise in a
single 0.66~\kms\ wide channel is 20~\mjb\ toward the NE pointing and
15~\mjb\ toward the SW pointing.  These noise levels are about five
times better than those reported for the ensemble of sources observed by
\citet{wang}.  The synthesized beam for these maser data is approximately
$1\farcs5 \times 0\farcs9$.

\section{Results}
\subsection{NH$_3$}

\begin{figure}[hbtp]
   \epsscale{1.0}
    \plotone{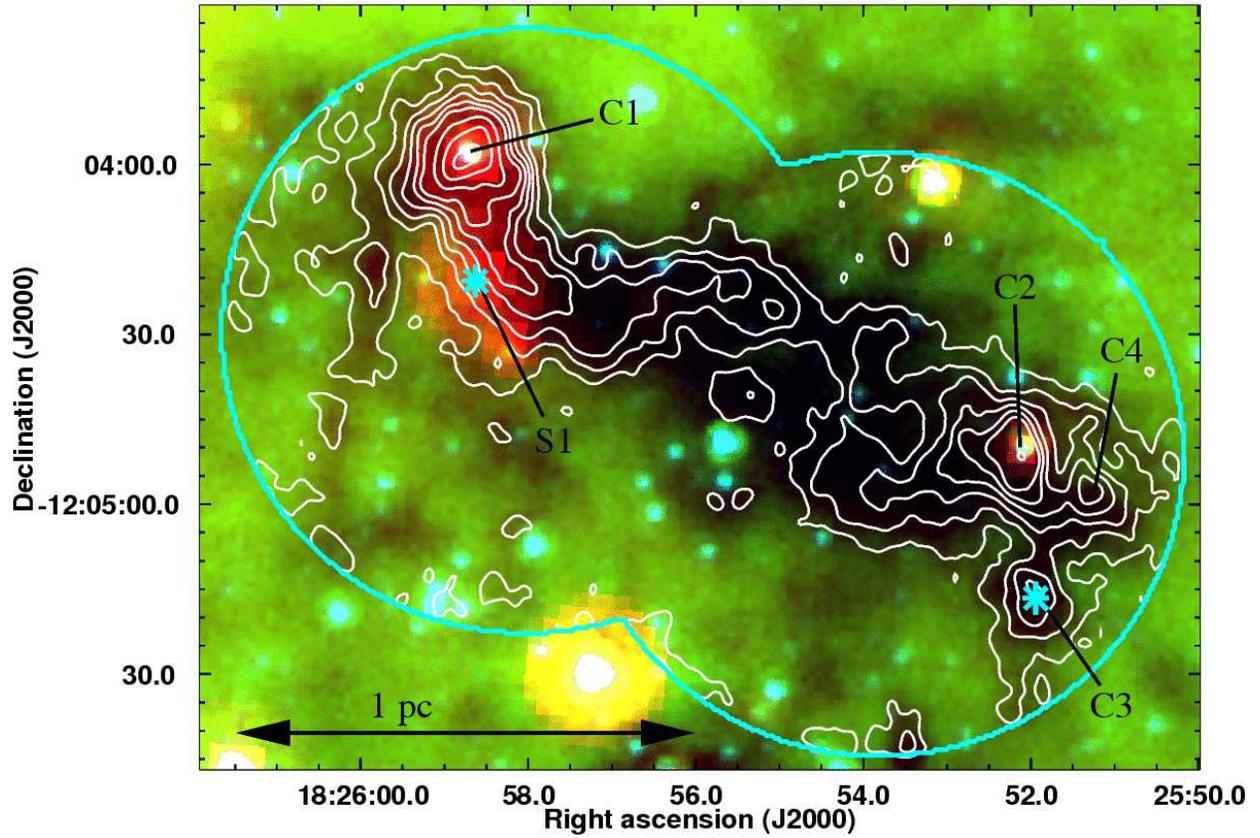}
\caption{\footnotesize 4.5~\mum\ (blue), 8.0~\mum\ (green), and 24~\mum\
(red) 3-color image of IRDC G19.30+0.07.  White contours represent the
integrated intensity from the main hyperfine component of the NH$_3$ (1,1)
line.  The NH$_3$ (1,1) resolution is $4\farcs5 \times 3\farcs4$ and RMS
noise at the center of the primary beam is $\sim 5$~mJy~beam$^{-1}$~\kms.
Contours mark 4$\sigma$, 8$\sigma$, 12$\sigma$, 16$\sigma$, and 20$\sigma$
and then increase in 8$\sigma$ intervals, where $\sigma$ is defined at
the center of the primary beam.  Blue lines mark the 50\% power point of
the primary beam; noise levels at the edge of the field are twice those
at the center of the primary beam.  The light blue asterisks indicate
H$_2$O maser detections \citep[][this paper]{wang}.  The four NH$_3$
clumps identified by this work are labeled C1, C2, C3, and C4.}
\label{3col19}
\end{figure}

The NH$_3$ (1,1) main line emission, integrated over 24.1 to 30.3~\kms,
is shown overlaid on an infrared 3-color Spitzer image of G19.30+0.07
in Fig.~\ref{3col19}.  The three colors in Fig.~\ref{3col19} are GLIMPSE
4.5~\mum\ (blue) and 8~\mum\ (green) emission and MIPSGAL \citep{mipsgal}
24~\mum\ emission (red).  Fig.~\ref{3col19} shows that the NH$_3$ (1,1)
emission is spatially well correlated with the 8~\mum\ extinction.
NH$_3$ (2,2) emission was also detected throughout the IRDC, although
due to the fainter line emission the (2,2) emission was not as well
defined near the IRDC boundaries.  The RMS noise in the NH$_3$ (1,1)
integrated intensity image is 5~mJy~beam$^{-1}$~\kms.

In our discussions of IRDC structure, we use the guidelines described
by \citet{bergin}: {\it clouds} have masses of $\sim 10^3$ to
10$^4$~M$_{\sun}$ and radii of $\sim 1$ to 10~pc, {\it clumps} have
masses of $\sim 10^2$ to 10$^3$~M$_{\sun}$ and radii of $\sim 0.1$ to
1~pc, and {\it cores} have masses of $\sim 1$ to 10$^2$~M$_{\sun}$ and
radii of $\sim 0.01$ to 0.1~pc.  Clumps produce star clusters, and cores
produce individual stars.  A 20\arcsec\ FWHM Gaussian unsharp mask was
used to identify the compact structure in G19.30+0.07.  The selection
criteria for NH$_3$ clumps were peaks having structure $\la 20\arcsec$
with integrated intensities $\ga 25$~Jy~beam$^{-1}$~\kms\ in the masked
image.  G19.30+0.07 contains four apparent NH$_3$ clumps, labeled
in Fig.~\ref{3col19} as C1, C2, C3, and C4.  C1 and C2 coincide with
24~\mum\ emission and have previously been studied at mm wavelengths with
$\sim 11\arcsec$ resolution \citep{rathborne06}.  C3 is not associated
with any 24~\mum\ emission; however, a H$_2$O maser is located nearby
(see \S3.4).  C4 is an NH$_3$ emission peak not associated with any
IR emission or masers.  The 24~\mum\ source S1 (Fig.~\ref{3col19}) is
not coincident with a peak in the NH$_3$ emission.  There is a H$_2$O
maser with velocity 19~\kms\ \citep{wang} spatially coincident with S1.
This maser's velocity is slightly offset from the velocity of G19.30+0.07,
$\sim 26$~\kms.  While it is possible that the maser is associated with
an unrelated object along the line of sight, the probability of two
dense cores capable of forming a H$_2$O maser along the line of sight
to S1 is small.  It is more probable that the velocity offset between
the maser and the IRDC is caused by motion within the IRDC related to
star formation activity \citep{wang}.

\begin{figure}[hbtp]
    \epsscale{1.0}
    \plotone{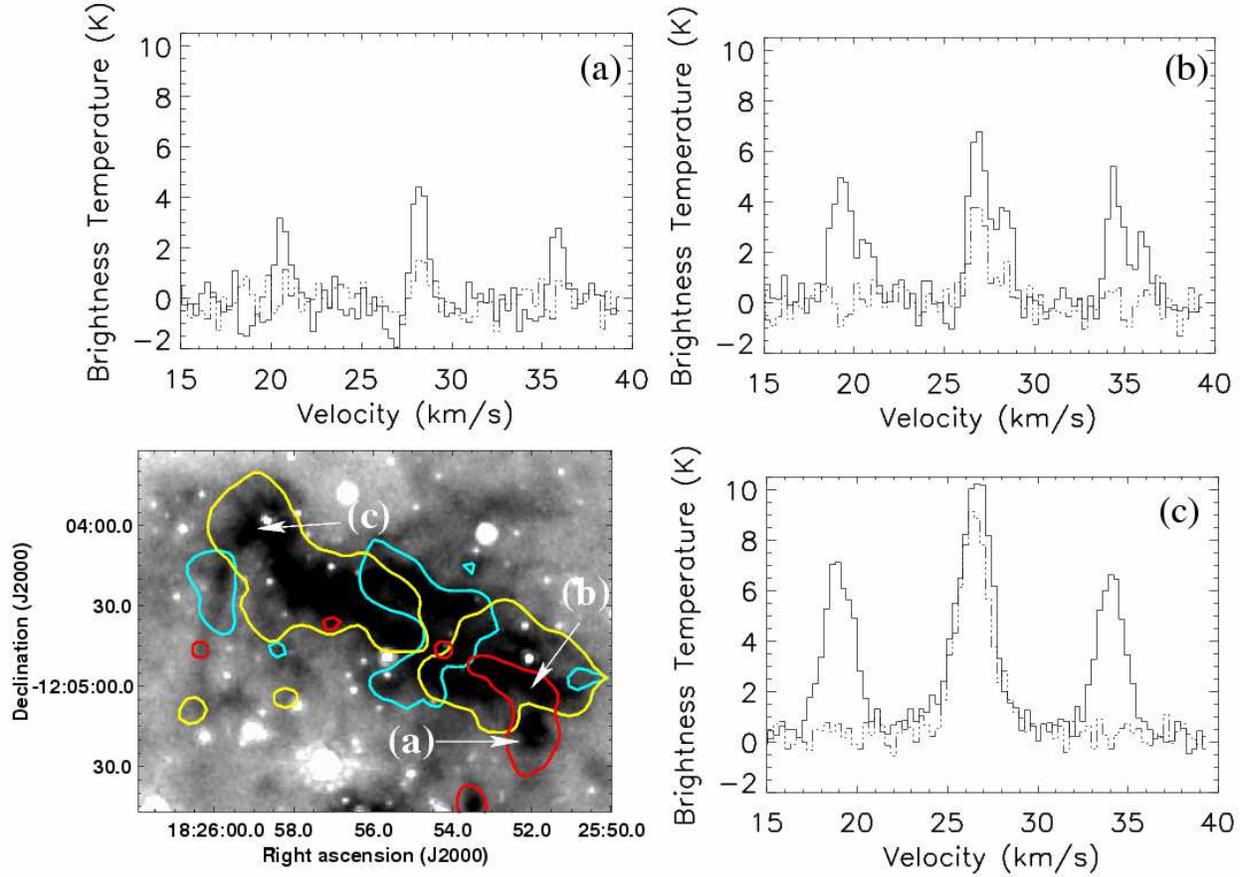}
\caption{\footnotesize Three representative line profiles of
NH$_3$ (1,1) and (2,2) emission from three different positions in
G19.30+0.07.  The solid and dot-dashed lines show the (1,1) and
(2,2) data, respectively.  Profile (a) shows a single velocity
line profile, observed at $\alpha$(J2000)=18$^h$25$^m$52.2$^s$,
$\delta$(J2000)=$-$12$^{\circ}$05\arcmin20\farcs7.  Profile (b)
shows emission from multiple velocity components, separated by
$\sim 1.5$~\kms, observed at $\alpha$(J2000)=18$^h$25$^m$52.1$^s$,
$\delta$(J2000)=$-$12$^{\circ}$04\arcmin59\farcs9.  Profile (c) is from a
point in the northern ammonia clump, $\alpha$(J2000)=18$^h$25$^m$58.8$^s$,
$\delta$(J2000)=$-$12$^{\circ}$04\arcmin02\farcs9, showing a broad
line-wing component.  The locations of the representative profiles are
indicated on the GLIMPSE 8~\mum\ image.  The line profiles were used to
identify three subclouds, separated by velocity offsets $\sim 1$~\kms.
The subcloud locations are shown overlaid on the GLIMPSE 8~\mum\ image;
the subcloud velocities are centered at 28.4~\kms\ (red), 26.7~\kms\
(yellow), and 25.7~\kms\ (blue).}
  \label{profile}
\end{figure}

Multiple velocity components and broad line-wing structures are
evident in the NH$_3$ emission, as shown in Figure~\ref{profile}.
Double-line profiles indicate locations in the cloud where there are two
overlapping kinematic components.  The double-line profile features cannot
be attributed to self-absorption: as demonstrated in Fig.~\ref{profile}b,
the double-line profile structure is seen in the more optically thin (2,2)
emission as well as the (1,1) emission.  Furthermore, the double-line
profiles are not correlated with NH$_3$ high optical depth (see \S4.2).
Finally, the individual components of the double-line profiles can
be traced continuously from regions with single-line profiles at the
same velocity; it is in overlap regions that they appear double, as
illustrated by the profiles shown in Fig.~\ref{profile}.  There are three
distinct velocity components over the entire G19.30+0.07 dark cloud,
which we will henceforth refer to as ``subclouds."  The spatial extents
of these subclouds are derived from fits to the NH$_3$ line profiles,
as described in \S4.1.  The subcloud positions relative to the 8~\mum\
emission are shown in Fig.~\ref{profile}.

\subsection{CCS}

\begin{figure}[hbtp]
   \epsscale{1.0}
    \plotone{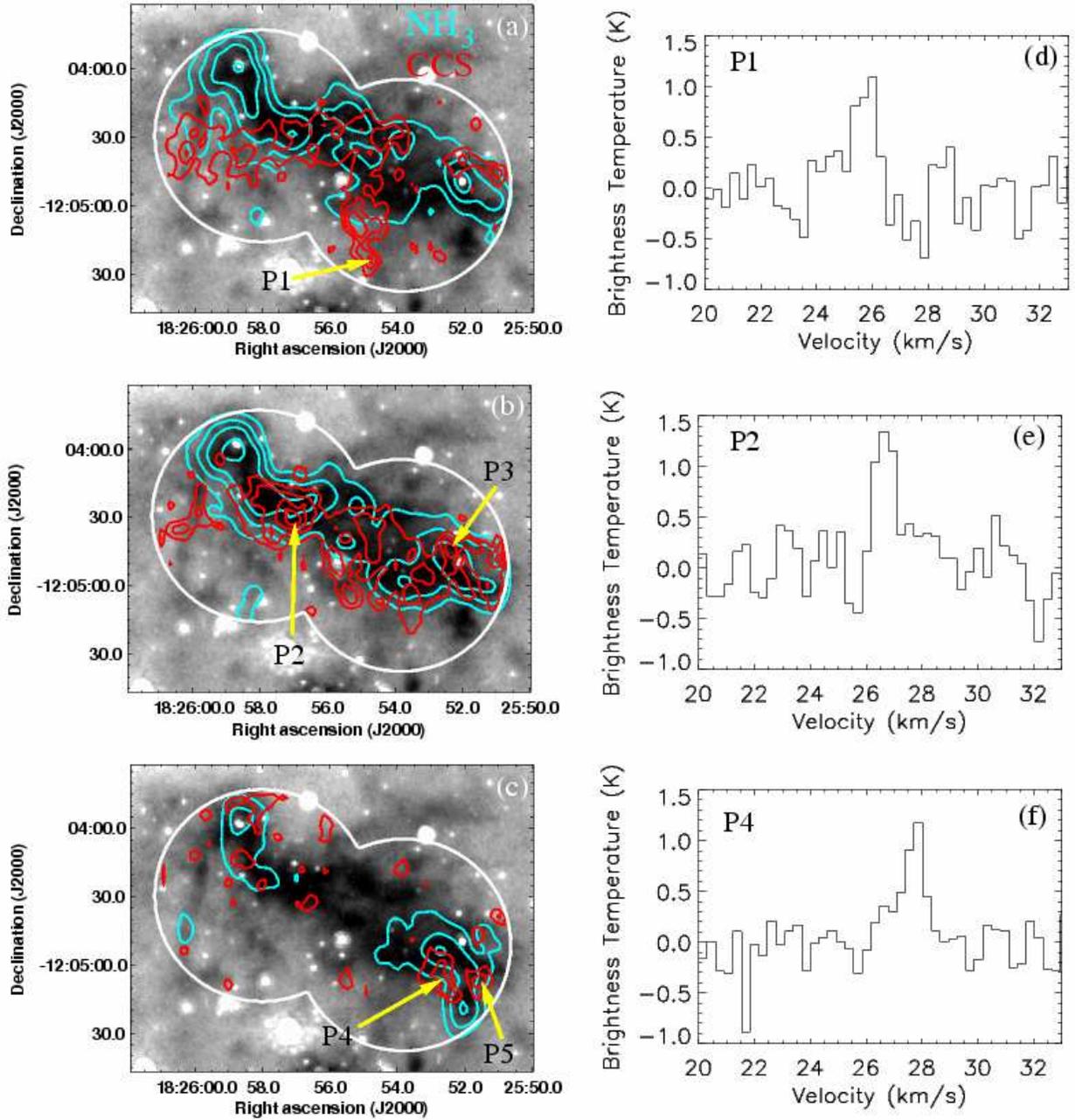}
\caption{\footnotesize CCS ($2_1-1_0$) distribution in IRDC G19.30+0.07
and representative profiles of CCS emission.  The greyscale is the
8.0~\mum\ image from GLIMPSE.  Red contours represent CCS ($2_1-1_0$)
emission integrated over different velocity components: (a) 25.2
to 26.5~\kms, (b) 26.2 to 27.4~\kms, and (c) 27.7 to 28.9~\kms.
The angular resolution of the CCS image is $7\farcs7 \times 5\farcs7$.
The CCS RMS noise is 1.3~\mjb~\kms.  CCS increments start at 3$\sigma$
and increase in 2$\sigma$ increments, where $\sigma$ is defined at
the center of the primary beam.  The white contours show where the
detection power of the primary beam in the CCS observations drops
to 65\%.  Blue contours represent NH$_3$ (1,1) emission smoothed to
match the CCS resolution (7\farcs7$\times$5\farcs7) and averaged over
the same velocity ranges.  The NH$_3$ RMS noise is 1.8~\mjb~\kms, and the
contours indicate emission levels of 10$\sigma$, 30$\sigma$, 60$\sigma$,
and 100$\sigma$.  Arrows indicate CCS peaks, labeled P1, P2, P3, P4,
and P5. Representative profiles toward P1, P2, and P4 are shown in (d),
(e), and (f), respectively.} \label{ccs}
\end{figure}

Where both CCS and NH$_3$ emission is detected their velocities match
very well.  However, in many locations there is emission from one molecule
and not the other.  Figure~\ref{ccs} shows the distribution of the
CCS and the NH$_3$ (1,1) main line emission along with representative
CCS profiles; peaks in the CCS emission have been labeled as P1-P5.
The CCS in Fig.~\ref{ccs} is integrated over four channels (1.2~\kms)
centered on the velocities of the subclouds as determined from the NH$_3$
emission.  The NH$_3$ emission in Figure~\ref{ccs} has been smoothed to
match the resolution of the CCS emission ($7\farcs7 \times 5\farcs7$)
and integrated over the same velocity ranges.  The RMS noise in the CCS
integrated intensity images is 1.3~mJy~beam$^{-1}$~\kms.

\subsection{8.4 GHz Continuum}

No 8.4~GHz continuum emission is detected from G19.30+0.07.  The RMS
noise in the field is 0.025~mJy~beam$^{-1}$.  A 5$\sigma$ (0.125~mJy)
upper limit on any free-free continuum emission implies an upper limit on
the UV photon flux of $N_c \la 8 \times 10^{43}$~s$^{-1}$ \citep{mezger,
condon}.  However, it is unlikely that all of the emitted UV photons go
into ionization of H and He.  The UV flux could be reduced by absorption
by an accretion disk or dust surrounding the protostar.  Thus, the UV
flux limit is likely an underestimate of $N_c$.  Furthermore, early-stage
\HII\ regions can be small, dense, and optically thick, and consequently
have low radio continuum flux densities.  Further discussion of the
protostellar mass limit implied by the UV flux limit is given in \S5.3.

\subsection{H$_2$O Masers}

From our re-reduction and analysis of the \citet{wang} H$_2$O
maser data we have detected two regions of maser emission.
A previously unreported maser with a peak flux density of
1.58~Jy was detected at $\alpha$(J2000)=18$^h$25$^m$51.95$^s$,
$\delta$(J2000)=$-$12$^\circ$05\arcmin14\farcs0 at an LSR
velocity of 28.8~\kms.  This maser is located within the C3
ammonia clump (Fig.~\ref{3col19}), and is the first reported
evidence of star formation associated with this clump.
The second H$_2$O maser, first reported in \citet{wang},
has a peak flux density and velocity of 0.51~Jy and 19.0~\kms,
respectively, at a position of $\alpha$(J2000)=18$^h$25$^m$58.64$^s$,
$\delta$(J2000)=$-$12$^\circ$04\arcmin21\farcs5.  This maser is located
close to the 24~\mum\ source S1.  The reported peak flux densities have
been corrected for primary beam attenuation.

\section{Physical Properties of the Dense Gas}

\subsection{Gas Kinematics}

The line profiles from the spatially smoothed $6\arcsec \times 6\arcsec$
resolution NH$_3$ (1,1) and (2,2) images have been fitted using a
non-linear least squares Gaussian fitting routine, providing central
velocities, linewidths, and amplitudes.  Fitting errors for the NH$_3$
central velocity and linewidths are typically $\la 0.1$~\kms, and $\la
10$\% for the amplitude, with the uncertainty reduced by approximately
a factor of two in areas with a high signal-to-noise ratio.

The (1,1) line fitting was done in two steps.  First, the main (1,1) line
was fitted.  Second, the main line and two satellite hyperfine lines were
fit simultaneously with a triple-gaussian profile, using the results from
the first (main-line only) fit to set initial guesses for, and constraints
on, the fits to the satellite lines.  The initial guesses and constraints
were set as follows: (1) the initial guesses for the central velocities
of the satellite lines were set at the velocity of the main line $\pm
7.7$~\kms, constrained to $\pm 0.6$~km/s of this initial guess; (2)
the initial guess for the width of the satellite lines was set to the
width of the main line, constrained to $\pm 0.6$~\kms.  The fit to the
(2,2) line was not constrained by the results of the (1,1) fit.

In areas with double-line profiles (see Fig.~\ref{profile}), the fitting
included additional steps.  One of the (1,1) main lines and its associated
satellite lines were fitted as described above.  Then the second (1,1)
main and satellite lines were fitted as described above.  Finally, the
results of these fits were used to set the initial guesses and constraints
for a simultaneous (six-gaussian) fit that separated the various emission
contributions.  The double-profile fits of the satellite lines used the
same main-line based constraints as described for single-profile fits.
The fits to (2,2) double-line profiles were done simultaneously using
a two-gaussian profile; as in the single-profile case, the fits were
not constrained by the (1,1) line fits.  C1 and C2 exhibit an emission
component with broad line-wings, possibly indicating outflows; in these
areas the line-wing component was modeled as an additional Gaussian
during simultaneous line fitting.

\begin{figure}[hbtp]
  \vspace{9pt}
	\epsscale{1.0}
    \plotone{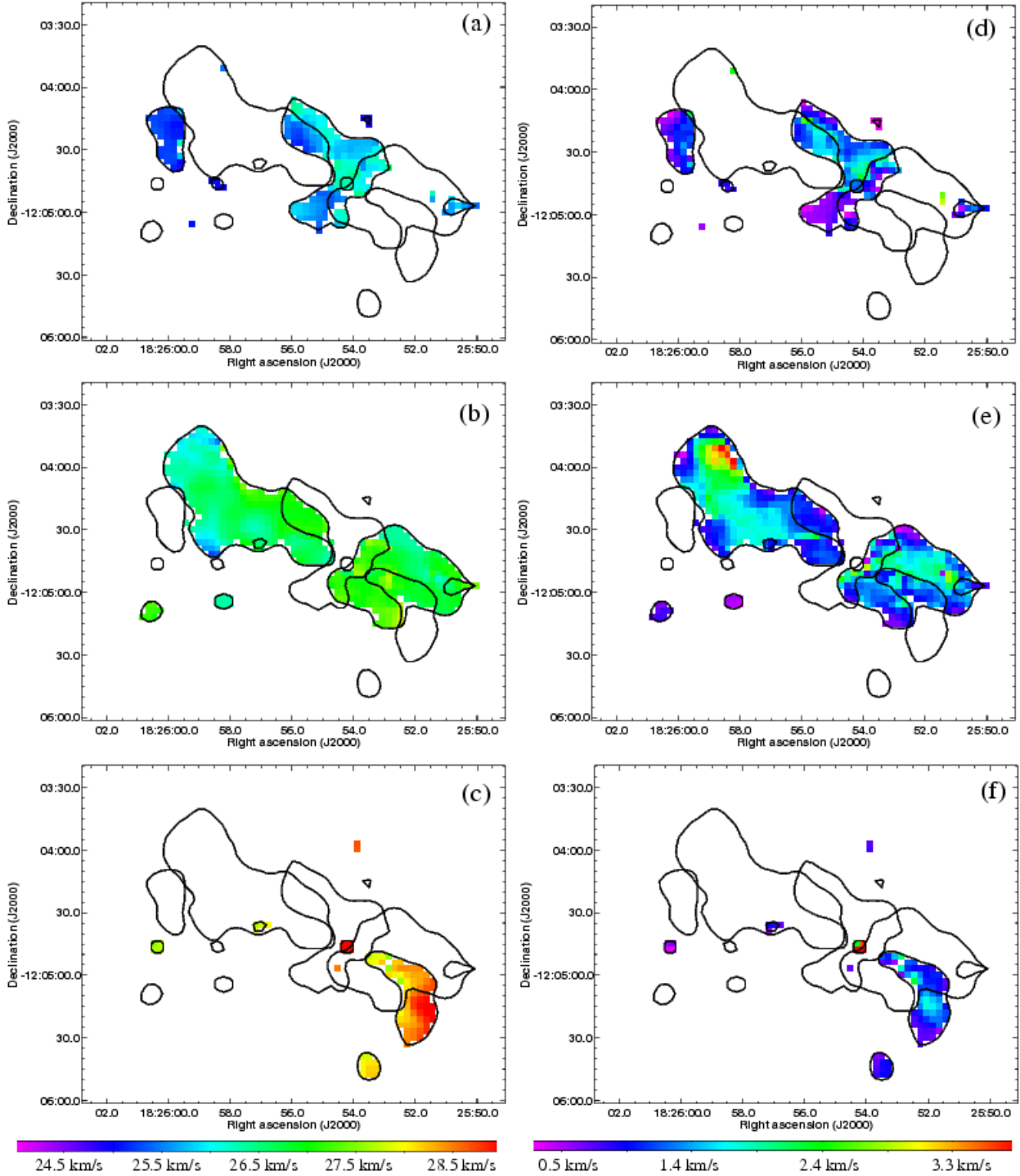}
\caption{\footnotesize Velocities and FWHM linewidths within the three
subclouds that make up G19.30+0.07.  Pixel spacing is 3\arcsec, resolution
is 6\arcsec.  Black outlines indicate the boundaries of the different
subclouds identified in Fig.~\ref{profile}.  The LSR velocities are shown
in (a), (b) and (c), and the NH$_3$ (1,1,$m$) FWHM linewidths are shown
in (d), (e), and (f).  Velocity components are centered at 25.7~\kms\
in (a) and (d), 26.7~\kms\ in (b) and (e), and 28.4~\kms\ in (c) and (f).}
  \label{velcombo}
\end{figure}

Three subclouds were identified based on fits of the NH$_3$ (1,1)
profiles.  Their median LSR velocities are centered at 25.7~\kms,
26.7~\kms, and 28.4~\kms.  The spatial distributions of the velocities and
linewidths for each of the subclouds are shown in Figure~\ref{velcombo}.
Internal motion is indicated within each subcloud by central velocity
changes of $\la 1$~\kms, shown in Figs.~\ref{velcombo}a, \ref{velcombo}b,
and \ref{velcombo}c.  Typical FWHM linewidths are $\sim 1.2$~\kms,
but the linewidths are $\sim 0.6$~\kms\ at the subcloud boundaries and
increase to $\ga 2$~\kms\ near C1 and C4, shown in Figs.~\ref{velcombo}d,
\ref{velcombo}e, and \ref{velcombo}f.  Since C4 contains no evidence of
star formation, the increased velocity dispersion at C4 may be due to
a nearby outflow originating from C2.

As discussed in \S3.2, CCS emission is found in all of the subclouds.
The CCS emission lines could not be fitted reliably because of their low
signal-to-noise ratio, but based on the profiles shown in Fig.~\ref{ccs}
we estimate the CCS FWHM to be $\sim 0.6$~\kms, and therefore assume the
CCS emission from each subcloud is contained within $\sim 1.2$~\kms\
centered at that subcloud's velocity.  Where both CCS and NH$_3$ are
detected, the linewidths and emission line structure from both molecules
are in good agreement, even in areas with multiple velocity components.
Thus, CCS velocity components were separated by averaging over 1.2~\kms\
centered at the subcloud velocities determined from NH$_3$.

\begin{figure}[hbtp]
    \epsscale{1.0}
    \plotone{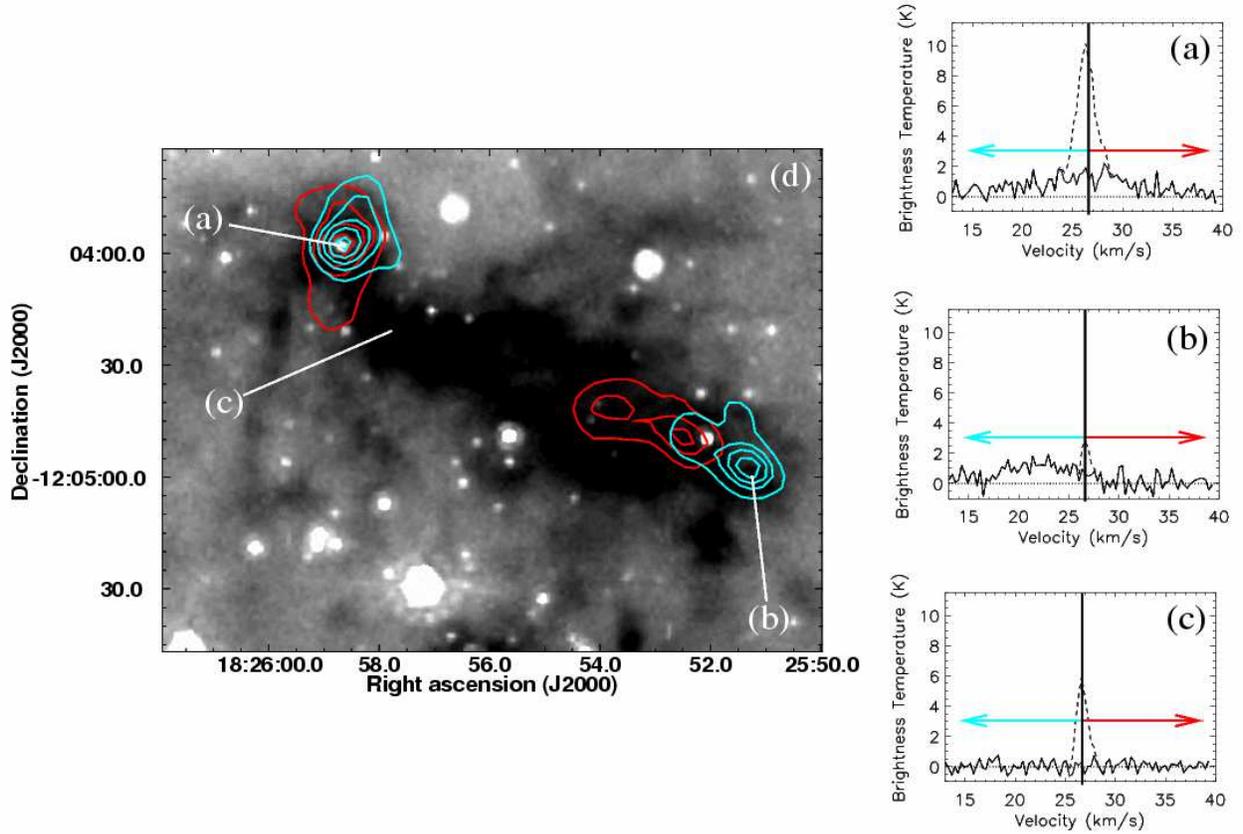}
\caption{\footnotesize Broad line-wings in addition to narrow NH$_3$
emission are found near C1 and C2.  Three representative profiles of the
NH$_3$ (2,2) emission are shown from different locations in G19.30+0.07.
In (a), (b), and (c), the total (2,2) line emission is shown as a dashed
line and the residual emission (line-subtracted, as described in the text)
is shown as a solid line.  In (a) and (b), broad line-wing components are
present; for comparison, (c) lacks line-wing emission.  The locations
of the representative profiles are indicated on the GLIMPSE 8~\mum\
image (d).  Contours in (d) show integrated intensity from the NH$_3$
(2,2) residual images.  The blue and red-shifted line-wing emission
was integrated over 14.6 to 26.6~\kms\ and 26.6 to 38.7~\kms; these
velocity ranges are indicated by red and blue arrows in (a), (b), and (c).
Blue contours represent the integrated intensity blue-ward of the NH$_3$
(2,2) line center, and red contours represent the integrated intensity
red-ward of the NH$_3$ (2,2) line center.  The 1$\sigma$ uncertainty in
the emission integrated over 6~\kms\ is $\sim 0.025$~\jb~\kms.  Contours
show 2$\sigma$, 6$\sigma$, and then increase in 3$\sigma$ intervals.}
  \label{outflow}
\end{figure}

NH$_3$ (2,2) line profiles were used to examine the kinematics of the
line-wing features, as the greater separation between the main line
and satellite hyperfine components for the (2,2) transition simplified
simultaneous fitting of narrow and broad components.  The line-wings
toward C1 and C2 had typical FWHM linewidths of $\sim 12$~\kms.
A residual image cube was made of the NH$_3$ emission with the
fitted narrow component subtracted, so that only the broad component
remained.  The integrated blue- and red-shifted emission was obtained
from the residual data by integrating over the velocity ranges 14.6
to 26.6~\kms\ and 26.6 to 38.7~\kms.  Representative profiles and the
red- and blue-shifted integrated NH$_3$ (2,2) emission are shown in
Figure~\ref{outflow}.

The red and blue line-wing emission from C1 is spatially coincident with,
and centered on, a 24~\mum\ source, and can be interpreted as an outflow
oriented close to the line of sight.  The infall velocity derived from
the mass and radius of C1 (see \S4.4) is $\sim 2$~\kms, insufficient to
account for the broadening observed in the line-wing emission.  The age
of C1's outflow cannot be determined from the flow velocities because
of the line-of-sight orientation.  The line-wing emission at C2 can be
interpreted as an outflow originating at the 24~\mum\ source, with the
red shifted outflow extending $\sim 0.4$~pc NE and the blue shifted
outflow extending $\sim 0.3$~pc SW of the source.  The outflow from
C2 appears well collimated, with a collimation factor (length/width)
$\ga 5$.  The outflow's inclination angle is unknown, so we assume an
angle of 45$^{\circ}$ in estimating the measured length of the outflow.
The central velocity derived from a Gaussian fit to the broad component
at each pixel within the C2 outflow gives an average outflow velocity of
$\sim 5$~\kms, corrected for an assumed inclination angle of 45$^{\circ}$.
The outflow's kinematic age was determined by dividing the extent of
the outflow by the average outflow velocity, resulting in an age of
$\sim 10^5$~yr.

\subsection{Optical Depth, Temperature, and Density Structure}

NH$_3$ inversion transitions have hyperfine splitting due to magnetic
dipole interactions and the interaction of the nitrogen electric
quadrupole moment and the electron electric field \citep{ho83}.
The optical depth of the NH$_3$ (1,1) main line, $\tau_{(1,1,m)}$,
can be derived from the ratio of the main and satellite hyperfine
line strengths.  Assuming the excitation temperature is the same for
the main and satellite lines, the ratio of the brightness temperatures
$T_{b(1,1,m)}$ and $T_{b(1,1,s)}$ can be expressed as

\begin{small}
\begin{equation}
\frac{T_{b(1,1,m)}}{T_{b(1,1,s)}}
     = \frac{1-\exp[-\tau_{(1,1,m)}]}{1-\exp[-\alpha\tau_{(1,1,m)}]}
\end{equation}
\end{small}

\noindent where `$m$' and `$s$' indicate the main and satellite hyperfine
components, $T_b$ is the measured brightness temperature, $\tau_{(1,1,m)}$
is the optical depth of the main quadrupole hyperfine component of the
(1,1) line, and $\alpha$ is the theoretical intensity ratio of the
satellite line to the main line if the hyperfine levels are populated
according to their statistical weights.  For the innermost satellite
hyperfine transition, $\alpha = 0.28$ \citep{rydbeck}.

The total optical depth of the (1,1) line, $\tau_{(1,1)}$, can be
determined from $\tau_{(1,1,m)}$ and the relative intensities of the
hyperfine components, such that $\tau_{(1,1)} = \zeta \tau_{(1,1,m)}$,
where $\zeta$ is the ratio of the total (1,1) line optical depth to
the optical depth of the main hyperfine component.  If the NH$_3$ (1,1)
hyperfine levels are populated according to their statistical weights,
$\zeta = 2$ \citep{rydbeck}.

\begin{figure}[hbtp]
    \epsscale{1.0}
    \plotone{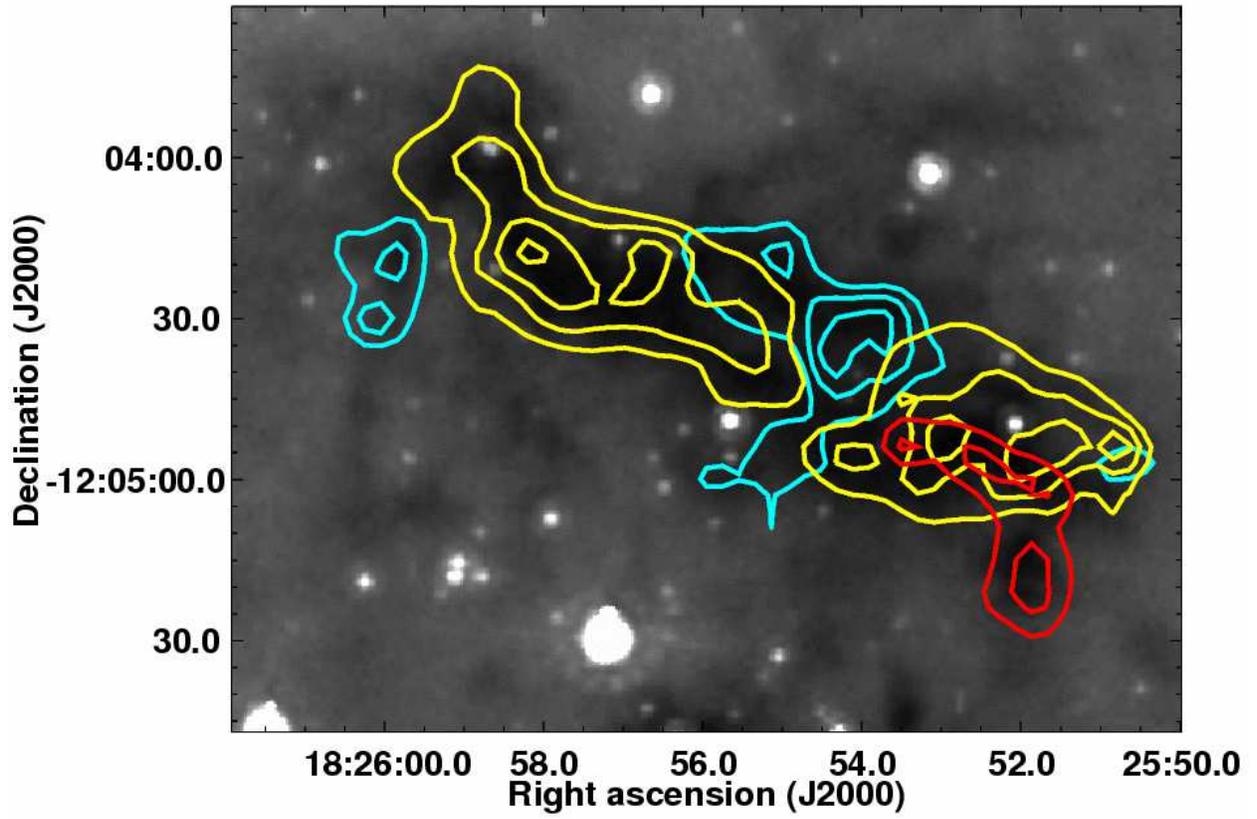}
\caption{\footnotesize The optical depth of NH$_3$ (1,1) in the
three subclouds is shown. The components are centered at 28.4~\kms\
(red), 26.7~\kms\ (yellow), and 25.7~\kms\ (blue).  Contours start at
$\tau$(1,1)=3 and increment by 3.}
  \label{tau}
\end{figure}

The optical depth $\tau_{(1,1)}$ ranges from $\sim 1$ to 4 at the
boundaries of the subclouds and increases towards the cloud centers.
Near C1, C2, and C3 $\tau_{(1,1)}$ is $\sim 5$ to 10.  The uncertainty in
$\tau_{(1,1)}$ is $\la 30$\% where the NH$_3$ signal to noise is highest.
At the subcloud boundaries, the uncertainty in $\tau_{(1,1)}$ increases,
and is on the order of a factor of two.  At positions where $\tau_{(1,1)}
> 20$, $T_{b(1,1,s)}$ approaches $T_{b(1,1,m)}$ too closely to yield
accurate results; at these positions a $\tau_{(1,1)}$ value of 20 was
assumed.  The optical depth distribution is shown in Figure~\ref{tau}.
It is notable that areas where NH$_3$ double-line profiles are detected,
shown as subcloud overlap in Fig.~\ref{profile}, are not coincident with
the highest optical depth gas, ruling out self-absorption as a cause of
the double peaked line profile.

The ratio of the brightness temperatures of the NH$_3$ (1,1,$m$)
and (2,2,$m$) lines can be used to determine the rotation temperature
$T_R(1,1:2,2)$ that characterizes the population distribution among the
(1,1) and (2,2) states \citep{ho83}:

\begin{small}
\begin{equation}
T_R(1,1:2,2)=-41.5
\left(
	\ln{\biggl[
		-\frac{0.284}{\tau_{(1,1,m)}} 
		\ln{\Bigl[
			1-\frac{T_{b(2,2,m)}}{T_{b(1,1,m)}}\bigl(
				1-\exp[-\tau_{(1,1,m)}]
			\bigr)
		 \Bigr]}
	\biggr]}
\right)^{-1}.
\end{equation}
\end{small}

\begin{figure}[hbtp]
    \epsscale{0.7}
    \plotone{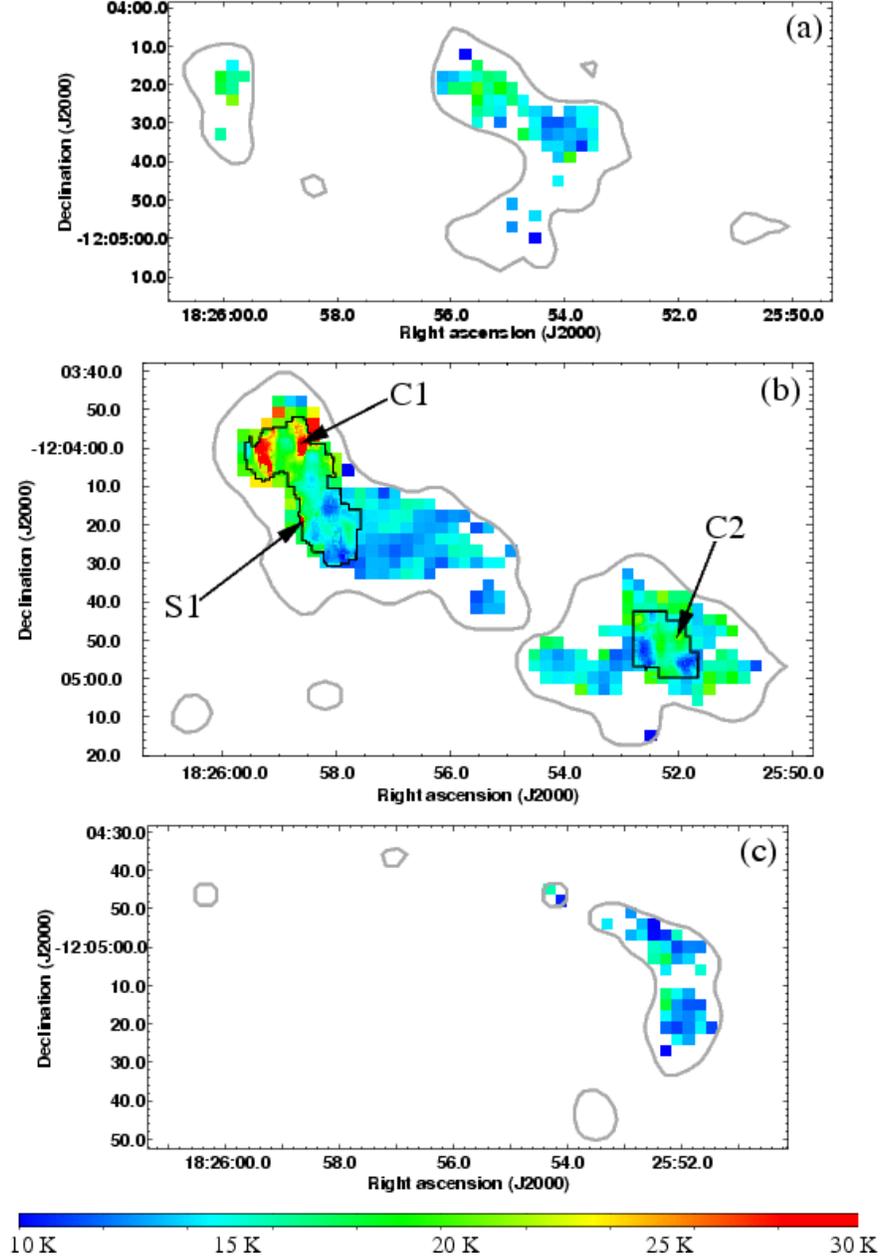}
\caption{\footnotesize The NH$_3$ rotation temperature distribution
in the three subclouds is shown.  Only points with uncertainty $\le
10$\% have been mapped.  Grey outlines indicate the boundaries of the
different subclouds.  The component centered at 25.7~\kms\ is shown in
(a), 26.7~\kms\ in (b), and 28.4~\kms\ in (c).  The resolution in the
images is 6\arcsec\ except in the areas outlined in black in (b), which
have a higher resolution of $4\farcs5 \times 3\farcs4$ and pixel spacing
0\farcs5.  Arrows in (b) indicate the locations of C1, C2, and S1.}
  \label{temperature}
\end{figure}

\noindent The distribution of $T_R$ for locations where the statistical
uncertainty in $T_R$ is 10\% or less is shown in Figure~\ref{temperature}.
The $T_R$ gradient is steep near the NH$_3$ clumps, and consequently
$T_R$, in particular, suffers from the effects of beam dilution caused by
smoothing the data.  Thus, we created temperature maps at the original
$\sim 4$\arcsec\ resolution of C1 and C2, where the NH$_3$ emission is
strong enough to obtain good line fits.  The higher resolution images
are shown outlined in black in Fig.~\ref{temperature}(a).

Values of $T_R$ range from 10~K up to $\sim 30$~K, above which
$T_{b(2,2,m)}$ approaches $T_{b(1,1,m)}$ too closely to yield reliable
temperature estimates.  A value of 35~K was assumed at points where
$T_{b(2,2,m)} \sim T_{b(1,1,m)}$ to indicate points with uncertain
temperatures.  The higher resolution images of the clumps show that
$T_R$ reaches this maximum at C1 and S1, and $T_R \sim 20$~K in C2,
indicating heat sources at these locations.  Heating is consistent with
the presence of 24~\mum\ emission in C1, C2, and S1.  An area with $T_R
\ge 35$~K approximately 10\arcsec\ east of C1 has no corresponding peak
in NH$_3$, and the nature of this peak in $T_R$ remains unknown.

Using earlier results from \citet{danby}, \citet{tafalla} empirically find
a relationship between IRDC gas kinetic temperature, $T_K$, and $T_R$,

\begin{small}
\begin{equation}
T_K \approx
  \frac{T_R}
  {1-\frac{T_R}{42}{\rm ln}\Bigl[1+1.1{\rm exp}(\frac{-16}{T_R})\Bigr]}.
\end{equation}
\end{small}

\noindent This relationship is good to better than 5\% when $T_R$ is 5
to 20~K.  Above 20~K, the relationship is not as accurate.  Since the
majority of the gas in G19.30+0.07 has $T_R \le 20$~K, we apply this
relationship and find a mean $T_K$ in the IRDC of $\sim 17$~K compared
to a mean $T_R$ of $\sim 15$~K.

The column density of the NH$_3$ (1,1) transition was calculated from the
total optical depth and brightness temperature of the (1,1) transition,
$\tau_{(1,1)}$ and $T_{b(1,1)}$, which include all hyperfine components.
In the optically thin limit, $T_{b(1,1)}$=$\zeta T_{b(1,1,m)}$, where
$\zeta = 2$ and is defined as above.  The column density of the NH$_3$
(1,1) transition was determined using

\begin{small}
\begin{equation}
N_{(1,1)} = T_{b(1,1)}
   \frac{\Delta V}{\sqrt{8\ln(2) }} 
   \frac{\tau_{(1,1)}}{\left(1-\exp[-\tau_{(1,1)}]\right)}
   \frac{8 k \pi^{1.5} \nu_{(1,1)}^2}{h c^3} \frac{1}{A_{(1,1)}},
\end{equation} 
\end{small}

\noindent where $\Delta V$ is the FWHM linewidth (in velocity units),
$\nu$ is the frequency of the (1,1) transition, and $A_{(1,1)}$ is
the Einstein coefficient for the sum of all hyperfine transitions.
Equation 4 contains an optical depth correction \citep{goldsmith} that
corrects for the opacity of the (1,1) emission.  The total column density
is obtained using the NH$_3$ partition function, $Q[T_R]$:

\begin{small}
\begin{equation}
N_{total} = \frac{N_{(1,1)} Q[T_{R}]}{g_{1,1}}
            \exp\biggl[\frac{E_{(1,1)}}{k T_{R}}\biggr].
\end{equation} 
\end{small}

\noindent Equation 5 assumes the same $T_R$ among all NH$_3$ metastable
levels, with negligible contribution from non-metastable levels
\citep{bachiller}.  $E_{(1,1)}$ is the energy above the ground state and
$g_{(1,1)}$ is the total statistical weight of the (1,1) inversion state.
Values for the partition function were derived from JPL line catalog
values calculated at 9.375~K, 18.75~K, and 37.50~K \citep{pickett} using
a linear interpolation.  Equations 4 and 5 assume that the emission
fills the synthesized beam, and neglect the microwave background.
In areas where the excitation temperatures approach the cosmic microwave
background temperature (2.7~K), Equation 5 may underestimate the column
density by roughly a factor of two \citep{bachiller}.   Corrections for
the opacity of the (1,1) emission were included in the calculation of
the NH$_3$ column densities; however, the column densities will be lower
limits in areas of very high optical depth ($\tau(1,1)~\gtrsim~20$),
since the optical depth cannot be measured accurately in these areas.
Fortunately, such high-opacity regions cover a very small area of the
IRDC, and are not associated with the clumps.

\begin{figure}[hbtp]
    \epsscale{1.0}
    \plotone{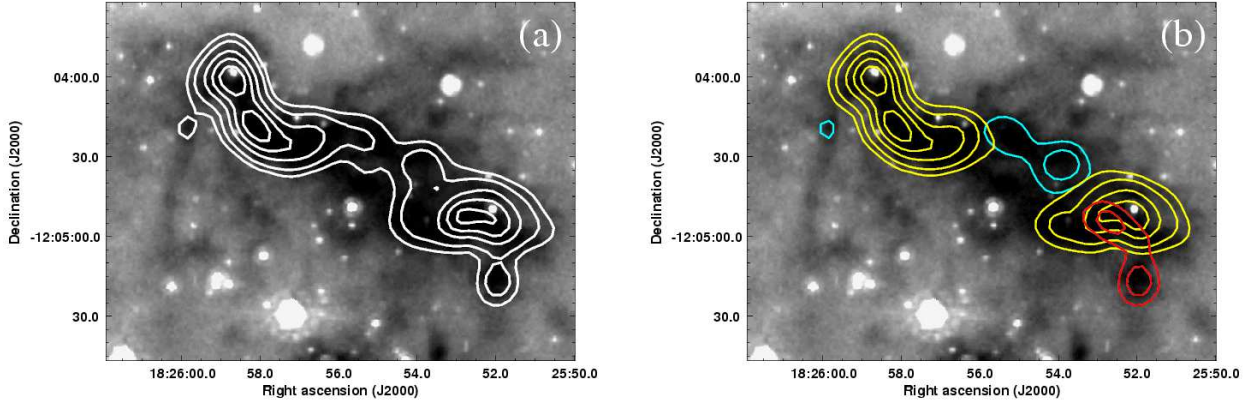}
\caption{\footnotesize The NH$_3$ column density integrated over all
velocities is shown in (a).  The column density contribution of the
separate velocity components is shown in (b).  The separated velocity
components are centered at 25.7~\kms\ (blue), 26.7~\kms\ (yellow),
and 28.4~\kms\ (red).  Contours indicate $10^{15}$~cm$^{-2}$, $2
\times 10^{15}$~cm$^{-2}$, and then increase in steps of $2 \times
10^{15}$~cm$^{-2}$.}
  \label{density}
\end{figure}

The NH$_3$ column density distribution is shown in Figure~\ref{density}.
The column densities lie in the range $10^{15}$ to $1.6 \times
10^{16}$~cm$^{-2}$.  In all three subclouds, there appears to be a
lower density envelope ($\sim 10^{15}$~cm$^{-2}$) around higher density
clumps ($\sim 4 \times 10^{15}$ to $1.6 \times 10^{16}$~cm$^{-2}$).
The uncertainty in column density is $\la 30$\% within the clumps,
but at the cloud boundaries the values are only good to approximately
a factor of two.  C1, C2, and C3 are coincident with column density peaks.

\citet{pillai} found the [NH$_3$]/[H$_2$] abundance ratio in IRDCs
ranged from $0.7 \times 10^{-8}$ to $10 \times 10^{-8}$.  We assume a
[NH$_3$]/[H$_2$] abundance ratio of $3 \times 10^{-8}$ \citep{harju}.
The H$_2$ column densities imply peak $A_V \gtrsim 100$ \citep{bohlin,
schultz}, in good agreement with the observed 8 \mum\ absorption.
To convert from column density to volume density, the line of sight
depth must be determined.  We assume a cylindrical geometry for the
filamentary IRDC, with the semi-major axis (``length") directed from the
northeast corner to the southwest corner of the IRDC.  We approximate
the semi-minor axis (``width'') of the IRDC, aligned perpendicular to
the semi-major axis, to be $\sim 0.35$~pc.  The assumption of cylindrical
geometry was then used to calculate the depth at each point in the IRDC.
The calculated H$_2$ number densities then range from $\sim 3 \times 10^4$
to $5 \times 10^5$~\cc, values consistent with previous number densities
reported for IRDCs \citep{carey98, egan98, pillai, wang08}.

\subsection{CCS Column Density}

The CCS ($2_1-1_0$) column density, $N_{1}$, is determined using 

\begin{small}
\begin{equation}
N_{1} = 
   \frac{8 \pi^{1.5} \nu^2 k}{h c^3} 
   \frac{g_1}{g_2} 
   \frac{1}{A_{21}} T_{b} 
   \frac{\Delta V}{\sqrt{8\ln(2) }},
\end{equation} 
\end{small}

\noindent where the symbols are defined as in Equations 4 and 5.
Equation 6 assumes the CCS ($2_1-1_0$) emission is optically thin
and fills the synthesized beam.  We assume $T_b \Delta V$ to be the
integrated emission of the CCS over 1.2~\kms.  Using the partition
function $Q[T_{ex}]$, where $T_{ex}$ is the temperature describing the
relative level populations, the total column density is obtained:

\begin{small}
\begin{equation}
N_{total} = \frac{N_{1} Q[T_{ex}]}{g_{1}}
            \exp\biggl[\frac{E}{k T_{ex}}\biggr].
\end{equation} 
\end{small}

\noindent where $E$ is the energy of the $1_0$ level above the ground
state.  Values for the partition function were interpolated from JPL
line catalog values in the same manner as for NH$_3$ \citep{pickett}.
We assume a CCS $T_{ex}$ of 10~K based on typical values of NH$_3$
$T_R$ in areas where CCS was detected.  Typical CCS column densities
ranged from $10^{13}$ to $3 \times 10^{13}$~\cmt, roughly two orders
of magnitude lower than the NH$_3$ column densities.  Unfortunately,
the unknown CCS $T_{ex}$ introduces significant uncertainty in the
column density measurements; using an excitation temperature of 5~K
reduces the column densities to $5 \times 10^{12}$ to $10^{13}$~\cmt,
while 15~K produces column densities ranging from $2 \times 10^{13}$
to $6 \times 10^{13}$~\cmt.  Due to the large uncertainties in the CCS
excitation temperature and linewidth, these CCS column densities are
likely accurate to only an order of magnitude.

The CCS $T_{ex}$ can be constrained using CCS (4$_3$--3$_2$) observations
of \citet{sakai08} (HPBW $\sim 37$\arcsec).  \citet{sakai08} used a
non-detection of CCS (4$_3$--3$_2$) emission to report an upper limit on
the total CCS column density of $\lesssim 3.7 \times 10^{12}$~\cmt\ at C1
and $\lesssim 3.5 \times 10^{12}$~\cmt\ at C2.  We used these results to
derive an upper limit for the CCS (4$_3$--3$_2$) column density, $N_{3}$,
at C1 and C2.  To compare our observations with those of \citet{sakai08}
we smoothed our CCS $N_{1}$ column density image to match the Nobeyama
Radio Observatory resolution, FWHM 37\arcsec, and measured $N_{1}$
at the \citet{sakai08} pointings.  By comparing our measured $N_{1}$
values and the $N_{3}$ upper limits and assuming local thermodynamic
equilibrium, we derived a CCS $T_{ex}$ upper limit of $\lesssim 5$~K.

\subsection{Mass}

IRDC and clump masses were derived from NH$_3$ column densities.
The mass of the IRDC was determined by integrating the NH$_3$ column
density (shown in Fig.~\ref{density}) over the entire source area
(clumps as well as diffuse gas), giving the total number of NH$_3$
molecules in the IRDC.  The number of molecules was converted to total
mass by assuming the mass is dominated by H$_2$ and using the assumed
[NH$_3$]/[H$_2$] abundance ratio of $3 \times 10^{-8}$ \citep{harju}
and an H$_2$ mass of $3.34 \times 10^{-24}$~g.  The IRDC mass is $\sim
1130$~M$_{\sun}$, consistent with the $\sim 900$~M$_{\sun}$ reported
by \citet{pillai}.  \citet{rathborne06} determined a total IRDC mass
of $\sim 400$~M$_{\sun}$ using 1.2~mm dust emission.  The apparent
discrepancy between the masses determined from NH$_3$ observations and
dust observations may be attributed to the large systematic uncertainties
in the dust emissivity, dust to gas ratio, and [NH$_3$]/[H$_2$] ratio.

The individual masses of C1, C2, C3, and C4 were calculated by integrating
NH$_3$ column density over only the clumps' spatial extents, and assuming
an [NH$_3$]/[H$_2$] abundance ratio of $3 \times 10^{-8}$.  The clump
radii were determined by fitting the NH$_3$ emission distribution in C1,
C2, C3 and C4 with a 2-D Gaussian using AIPS task JMFIT and using the
FWHMs as the radii of the clumps.  Clumps C1, C2, C3, and C4 have masses
140, 160, 30, and 30 M$_{\sun}$, respectively.  The mass uncertainties
given in Table~\ref{clumps} are based on uncertainty in the column
densities and clump radii; it should be noted that uncertainties in the
distance to the IRDC and the [NH$_3$]/[H$_2$] abundance ratio introduce
additional systematic uncertainty.  While the mass uncertainty may be
quite large, \citet{rathborne06} determined masses of 113~M$_{\sun}$
for C1 and 114~M$_{\sun}$ for C2 using 1.2 mm dust emission.
\citet{rathborne06} assumed a dust temperature of 15~K, comparable to
the temperatures we have measured in this IRDC (mean $T_K \sim 17$~K,
mean $T_R \sim 15$~K).  The clump radii determined by \citet{rathborne06}
were 0.15~pc for C1 and 0.21~pc for C2, comparable to the $\sim 0.1$~pc
determined for C1 and C2 in this study.  The \citet{rathborne06} masses
are in good agreement with those found in this study, despite the fact
that the overall IRDC mass determined by \citet{rathborne06} is a factor
of three lower than that determined in this study.  The fact that the
masses agree on the scale of the clumps but not on the scale of the
cloud may be due to sensitivity and/or resolution differences between
the two studies.

\section{Discussion}

\subsection{Clump stability}

We examined the stability of C1, C2, C3, and C4 by comparing their gas
mass to their virial masses.  The virial mass ($M_{vir}$) is the mass
at which thermal plus turbulent gas pressure balances gravitational
forces in the clumps, i.e., the gravitational and kinetic energy
are in equipartition.  The ratio of virial mass and clump mass
($M_{vir}/M_{clump}$) is a basic indicator of whether the clumps are
gravitationally bound and likely to collapse: if $M_{vir}/M_{clump}<1$,
the clump contains enough mass to collapse under gravity.  This balance
analysis assumes negligible contributions to clump stability from
additional forces such as magnetic fields and clump rotation, neither
of which are known for the clumps in G19.30+0.07.  Additionally, clump
confinement by external pressure \citep{bertoldi} has not been included.
We calculated $M_{vir}$ from the clump radius ($R_{clump}$, determined
in \S4.4) and the NH$_3$ FWHM linewidth ($\Delta V$) using

\begin{small}
\begin{equation}
M_{vir}=\frac{15 \Delta V^2 R_{clump}}{8 {\rm ln}(2) G} 
\end{equation} 
\end{small}

\noindent The NH$_3$ (1,1) FWHM linewidth was measured at the density peak
of the clumps.  Values of $\Delta V$, $M_{vir}$, and $M_{vir}/M_{clump}$
are listed in Table~\ref{clumps}.

The clumps all have $M_{vir}/M_{clump} > 1$, inconsistent with indications
of star formation that imply the clumps are undergoing gravitational
collapse.  However, Equation 8 assumes virial equilibrium.  If the
clumps have already collapsed and harbor outflows the conditions of
virial equilibrium are violated, and the linewidths will not accurately
predict the thermal and turbulent motion supporting the clumps.  Indeed,
clumps C1, C2, and C3 show evidence of outflow through either broad
NH$_3$ linewidths or H$_2$O masers.  Clump C4 has $M_{vir}/M_{clump}
\sim 8$; this high ratio may be a result of increased linewidths toward
C4 caused by the outflow originating from C2 (see \S4.1).  It is also
worth noting that where the cores are optically thick the $\Delta V$
may be overestimated, which would lead to a greater overestimate of
virial mass (which is proportional to $\Delta V^2$) than of clump mass
(which is proportional to $\Delta V$).

The virial masses of C1 and C2 reported by \citet{pillai} are 893
and 823 $M_{\sun}$, significantly larger than our $M_{vir}$ values.
\citet{pillai} used larger clump radii and higher linewidths in their
virial mass calculations because of the lower resolution of their study
(40\arcsec).  Our higher resolution shows most of the mass is contained
within a smaller radius than previously thought.  Furthermore, we were
able to resolve velocity gradients in the emission near the clumps
(Fig.~\ref{velcombo}).  The larger beam used by \citet{pillai} would
have confused this emission, contributing to their larger linewidth
measurements.

\subsection{The Relationship Between NH$_3$ and CCS}

Chemical evolution models comparing CCS and NH$_3$ distributions
have mainly been developed from observations of low mass star forming
regions and the environs of \HII\ regions; we now examine whether such
models might also apply in IRDCs.  Chemical models such as those of
\citet{suzuki}, \citet{bergin97}, and \citet{vandish} predict that
CCS traces chemically young gas in lower density, quiescent envelopes
of molecular clouds, while NH$_3$ traces dense, evolved clumps.
Time-dependent gas-phase modeling suggests that for low-mass star-forming
clumps the [CCS]/[NH$_3$] abundance ratio is a measure of the age of
clumps, i.e., a ``chemical clock'':  CCS abundance peaks in chemically
young gas and decreases as the CCS components are reprocessed into
molecules with stronger binding energies, such as CO.  \citet{millar}
and \citet{nejad} propose CCS is reprocessed on timescales of $\sim
10^5$~years, while de Gregorio-Monsalvo et al. (2005, 2006) concluded
that CCS has a lifetime of $\sim 10^{4}$ years after the onset of star
formation.

Freeze-out onto dust grains likely plays a significant role in chemical
changes during clump evolution.  Dust grain chemistry is particularly
important for explaining the high abundances of NH$_3$ observed in
dense clumps \citep{bergin97, bergin, sakai08}.  When CO freezes out
onto dust grains the abundance of NH$_3$ is enhanced, because gas-phase
CO destroys NH$_3$ parent molecules \citep{bergin97}.  Furthermore, the
abundance of hydrogenated molecules, including NH$_3$, may be enhanced
as ices evaporate off dust grains \citep{vandish} near hot clumps in
star-forming regions.  Freeze-out of CCS onto dust grains may also be
responsible for low CCS abundances in gas clumps where $n_{\rm H_2} >
10^6$~\cc\ \citep{bergin97}.  Dust grain chemistry also explains how the
chemical clock may be reset.  \citet{dickens} propose collisions in gas
with properties similar to IRDC subclouds can trigger MHD waves and/or
grain impacts that provide sufficient energy for thermal desorption of
molecules from grain surfaces, leading to the presence of early-time
molecules like CCS.

The critical densities of the NH$_3$ (1,1) and CCS ($2_1-1_0$) emission
lines differ by almost two orders of magnitude, $\sim 2 \times 10^3$~\cc\
for NH$_3$ \citep{danby} and $\sim 10^5$~\cc\ for CCS \citep{wolkovitch},
which can complicate the interpretation of molecular distribution.
However, despite its higher critical density, we detect CCS emission in
the lower density envelopes around the high-density NH$_3$ clumps rather
than toward the center of the dense clumps, suggesting their spatial
distributions are the result of chemical evolution in the IRDC rather
than a density effect.

Previous studies of the chemistry in G19.30+0.07 have been limited by
poor resolution.  Prior to this investigation, \citet{pillai} observed
G19.30+0.07 in NH$_3$ emission using the Effelsberg 100~m telescope
with 40\arcsec\ resolution, but lacked the resolution needed to comment
on small-scale structure.  \citet{sakai08} observed N$_2$H$^+$(1--0),
HC$_3$N(5--4), CCS(4$_3$--3$_2$), CH$_3$OH(7--6), and NH$_3$ (1,1),
(2,2), and (3,3) emission towards the clumps in G19.30+0.07 with the
NRO 45-m telescope (HPBW $\sim 37$\arcsec\ for CCS and $\sim 73$\arcsec\
for NH$_3$).  \citet{sakai08} report abundances suggesting the clumps in
G19.30+0.07 are more chemically evolved than low-mass starless cores.
The high spatial resolution data we present here improve upon these
findings by showing how morphology, temperature, and density structure
in the IRDC impact the distributions of the NH$_3$ and CCS.

Fig.~\ref{ccs} shows the integrated CCS emission in each of the subclouds
and compares it to the NH$_3$ integrated emission.  Although CCS
emission is associated with the subclouds at 25.7~\kms, 26.7~\kms,
and 28.4~\kms, the spatial distribution of the CCS and NH$_3$ emission
are not coincident.  Enhancements in the CCS with clump-like morphology
are labeled as P1 to P5 in Fig.~\ref{ccs}; no evidence of NH$_3$ clumps
exist at these locations.

\begin{figure}[hbtp]
    \epsscale{1.0}
    \plotone{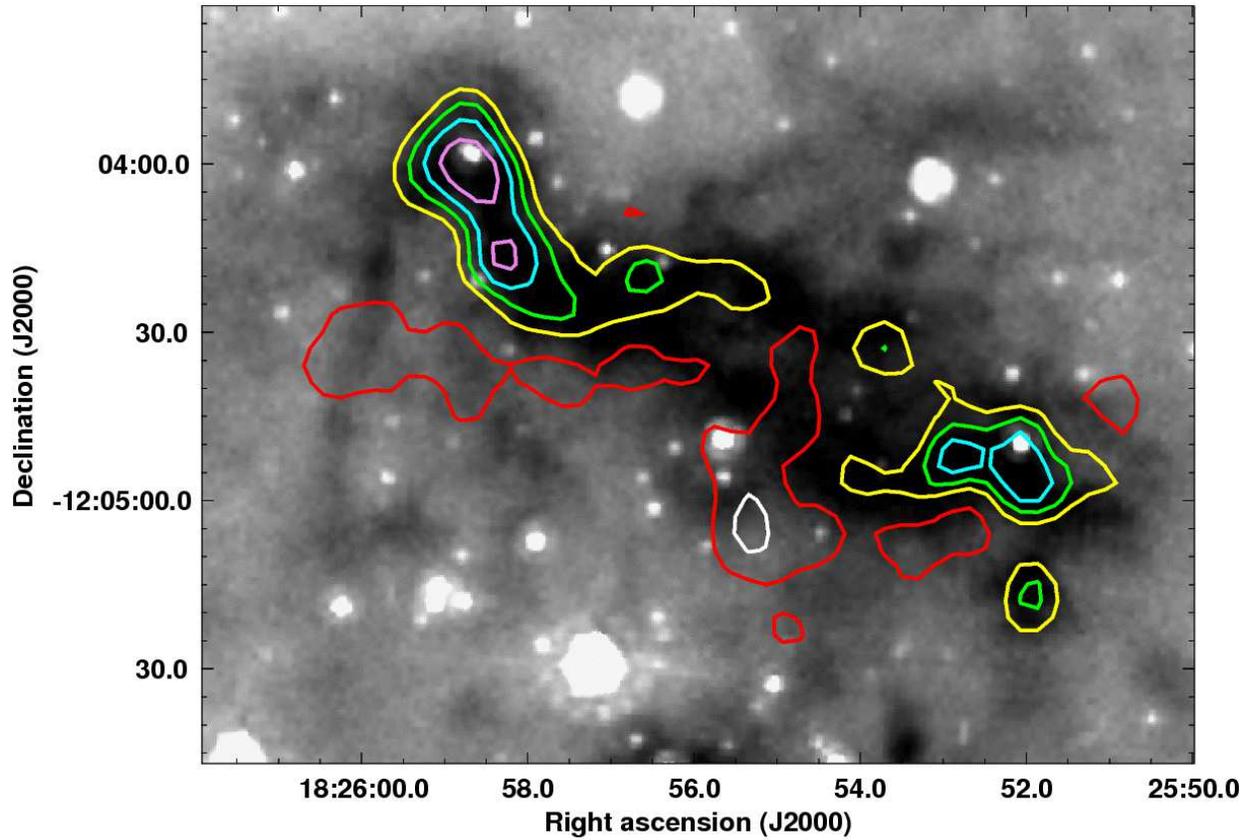}
\caption{\footnotesize Column density ratio map showing the relative
distribution of CCS and NH$_3$ ([CCS]/[NH$_3$]).  Only areas with
detectable emission from at least one of the molecules are mapped.
In CCS peaks lacking detectable NH$_3$ emission we assign an NH$_3$
column density upper limit of $10^{15}$~\cmt.  Where NH$_3$ but no CCS was
detected we assign a CCS column density upper limit of $10^{13}$~\cmt.
Contours scale logarithmically, with log([CCS]/[NH$_3$]) = $-$2.9
(violet), $-$2.7 (cyan), $-$2.5 (green), $-$2.3 (yellow), $-$1.9 (red),
and $-$1.7 (white).}
  \label{ratio}
\end{figure}

Typical CCS column densities ranged from $10^{13}$~\cmt\ to $3 \times
10^{13}$~\cmt.  We examined the [CCS]/[NH$_3$] column density ratio
in areas with detectable emission from at least one of the molecules.
In CCS peaks lacking detectable NH$_3$ emission we assign an NH$_3$
column density upper limit of $10^{15}$~\cmt.  Where NH$_3$ but
not CCS was detected we assign a  CCS column density upper limit
of $10^{13}$~\cmt.  At clumps C1, C2, C3, and C4 the [CCS]/[NH$_3$]
ratio is $\lesssim 10^{-3}$, slightly higher than the upper limit
determined by \citet{sakai08}, and consistent with active, star forming
clumps \citep{bergin, sakai08}.  At CCS emission peaks, [CCS]/[NH$_3$]
ratios are $\gtrsim 0.02$, consistent with results from \citet{suzuki},
who found [CCS]/[NH$_3$]~$\gtrsim 0.03$ in low-mass starless clumps.
A map of the column density ratios is shown in Figure \ref{ratio}.

Comparing Fig. \ref{ratio} with Fig. \ref{ccs} shows that P1 is associated
with the [CCS]/[NH$_3$] peak and located in the subcloud located at
25.7~\kms, the only subcloud lacking evidence of star formation.  The CCS
morphology and [CCS]/[NH$_3$] ratio in this subcloud are consistent
with a chemically young clump in a stage too early to present evidence
of star formation \citep{dgm06}.

The CCS peaks P2, P4, and P5 are located at subcloud interfaces, and P3
is spatially coincident with the outflow from C2.  The spatial coincidence
of these CCS peaks with subcloud-subcloud and outflow-subcloud interfaces
indicates collisions may be resetting the chemical clock of the gas at
these locations, consistent with the \citet{dickens} model.

One major caveat must be considered in the analysis of CCS emission.
CCS ($2_1-1_0$) is a low-lying energy state transition (upper energy level
only 1.6~K above ground), so it is difficult to know whether enhanced
CCS emission is due to excitation effects or increased abundance.
Using the results of \citet{sakai08} we were able to roughly constrain
the CCS excitation temperature to $\lesssim 5$~K.  However, the CCS
($4_3-3_2$) observations were low resolution and only provided a column
density upper limit.  Additional transitions of CCS must be measured at
comparable resolution to this study to more accurately determine the
CCS excitation temperature.  Sensitive interferometric observations at
millimeter wavelengths could explore whether the abundance measurements
in the ground state are consistent with abundances measured from higher
energy transitions.

We draw the following conclusions from the CCS detections: (1) CCS and
NH$_3$ are present in each of the three subclouds comprising G19.30+0.07,
but each molecule traces different parts of the subclouds; (2) CCS appears
to trace the boundaries of the IRDC while NH$_3$ peaks in the inner parts
of the IRDC, consistent with CCS enhancement in low-density envelopes
and NH$_3$ enhancement in denser clumps, in spite of the fact that the
CCS transition has the higher critical density.  This emphasizes that
NH$_3$ does not give the full picture of the dense gas in star-forming
regions; (3) in some cases, CCS appears to have clump-like morphology.
The [CCS]/[NH$_3$] peak at P1 is located in the only subcloud lacking
evidence of star formation.  This supports the hypothesis that CCS
cores may represent the very earliest stages of core formation; (4) CCS
peaks at subcloud-subcloud or outflow-subcloud interfaces may indicate
collisions are resetting the chemical clock in some parts of the IRDC;
and (5) further investigations at higher-energy transitions are needed
to distinguish excitation and abundance effects in CCS emission.

\subsection{Star Formation in G19.30+0.07}

Although early studies of G19.30+0.07 did not find any evidence of
star formation \citep{carey98}, recent investigations, including
this one, indicate substantial active star formation in this IRDC.
The observed evidence for star formation is: (1) archival MIPSGAL 24~\mum\
observations showing areas of hot dust associated with C1, C2, and S1;
(2) structure in the NH$_3$ spectral line observations consistent with
outflows originating from C1 and C2; (3) H$_2$O masers coincident with
C3 and S1 (Fig.~\ref{3col19}).

No 8.4~GHz continuum emission was detected in the IRDC.  A 5$\sigma$
detection limit of 0.125~mJy~beam$^{-1}$ sets the upper limit of the UV
photon flux at $< 8 \times 10^{43}$~s$^{-1}$, implying a stellar type
of B3 or later \citep{smith02, panagia}.  However, if the protostar is
accreting material, much of the stellar UV radiation will be absorbed
close to the star \citep{churchwell97}.  Also, in an IRDC it is likely
that some UV photons are absorbed by dust.  Thus, stars with types
hotter than B3 could be present in G19.30+0.07 but be undetected by our
measurements.  In fact, Orion Source I, an early-B star with mass $\ga
8$~M$_{\sun}$ \citep{matthews}, has an 8.4~GHz flux density of $\sim
1.1$~mJy \citep{menten} at a distance of $\sim 400$~pc \citep{menten2}.
If Orion Source I were located at the distance of G19.30+0.07, 2.4~pc,
its flux density would be only $\sim 0.03$~mJy, below the detection
limit of this study.

Clump masses of $\ge 10^2$~M$_{\sun}$ and the presence of outflows
suggest that C1 and C2 may be protoclusters.  Using the initial mass
function from \citet{scalo} and assuming a 50\% star forming efficiency,
these clumps would form stars up to masses of $\sim 3$~M$_{\sun}$.
The smaller NH$_3$ clump and H$_2$O maser without accompanying 24~\mum\
emission at C3 suggest C3 likely contains protostars that are either
lower-mass or younger than those in C1 and C2.  The 24~\mum\ emission
toward S1 suggests the presence of at least one protostar, consistent
with the H$_2$O maser coincident with S1.  It may be S1 is more evolved
than C1 and C2, and has destroyed or dispersed any dense gas associated
with its formation.  However, no shell of dispersed NH$_3$ is detected
around S1.

Assuming the linewing structure discussed in \S4.2 indicates outflows,
the outflow masses provide insight into the masses of the stars contained
in C1 and C2.  The column densities in the outflows were determined
as in Equations 6 and 7, using the integrated emission from the NH$_3$
(2,2) line-wings and assuming a $T_{ex}$ of 10~K for NH$_3$.  The column
densities were summed over the outflow area to determine the masses of the
outflows.  The red and blue outflow components of C1 each contain $\sim
2$~M$_{\sun}$ and the components in C2 each contain $\sim 1.6$~M$_{\sun}$,
suggesting intermediate or high mass protostar progenitors.  However,
we note that NH$_3$ only traces high density gas, so observations of the
low-density gas are needed to better constrain the total outflow masses.

Various models attempt to explain the formation of massive stars,
but recent work suggests accretion in turbulent clumps is the likely
creation mechanism \citep{beutherppv}.  \citet{beutherppv} present an
evolutionary sequence of massive stars starting from high mass starless
clumps: low- and intermediate-mass protostars form within the massive
starless clumps and accrete matter until becoming a massive protostellar
object, which develops into a massive star and clears the natal gas
from its environment.  C1 and C2 may be examples of clumps containing
the embedded accreting protostars proposed in this model.

Based on the results from this single IRDC, the role that subclouds play
in massive protostar and protocluster development is still unclear.  They
may indicate an early stage of fragmentation, cloud formation/collision,
or a superposition along the line of sight.  The subclouds appear
distinct, but if less dense gas below our detection limits connects the
subclouds, they may all be part of a larger structure.  Further work is
needed to understand the role of subclouds in IRDCs.

\subsection{Is G19.30+0.07 a ``Typical'' IRDC?}

Understanding the role of IRDCs in star formation requires a survey
of their properties.  We have examined one IRDC in great detail,
but is G19.30+0.07 a typical IRDC, and how general are our findings?
Our results are consistent with previous studies that indicate G19.30+0.07
has temperatures and densities typical for an IRDC \citep{carey98, pillai,
wang08, rathborne06}.  The high resolution of this study compared to
previous work has revealed new IRDC features.  Notably, the kinematics
have been analyzed in detail.  We find that the IRDC is composed of
three subclouds, each of which has a slightly different velocity.
The universality of multiple subclouds in IRDCs and the role they play
in protocluster evolution requires further investigation.

\section{Conclusions}

Massive star formation continues to be poorly understood, largely
because early stages are difficult to observe due to the short
timescales involved and the embedded nature of massive protostars.
Because of their large masses, low temperatures, and high densities,
IRDCs are currently the most likely candidates to host the first stages
of massive star formation.  Understanding their properties may shed
light on the processes that lead to the creation of massive protostars.
Although IRDCs are identified using IR observations, their high opacity
makes radio wavelengths an important probe of their structure.  We have
examined IRDC G19.30+0.07 using VLA spectral line observations of the
NH$_3$ (1,1), (2,2), and CCS ($2_1-1_0$) transitions.  From this work,
we draw the following conclusions:

\begin{enumerate}

\item G19.30+0.07 has structure down to the resolution limit of our
observations (4\arcsec\ to 6\arcsec).  At a distance of 2.4~kpc, the
structure is $\lesssim 1.4 \times 10^4$~AU in size.  This level of detail
is seen in the IR images as well as the NH$_3$ and CCS observations.

\item G19.30+0.07 is composed of three subclouds at different velocities.
Star formation and [CCS]/[NH$_3$] levels may indicate that one of the
subclouds is at an earlier evolutionary stage than the other two.  Further
investigation is needed to determine if IRDCs are typically composed of
multiple subclouds, and the role subclouds play in IRDC evolution.

\item Typical values of $T_R$ in G19.30+0.07 range from 10 to 30~K with
a mean value of $15 \pm 3$~K\@.  Increases in $T_R$ at C1, C2, and S1
indicate heat sources at these locations.

\item NH$_3$ column densities in G19.30+0.07 are $\sim 10^{15}$ to $1.6
\times 10^{16}$~cm$^{-2}$, with lower density envelopes around higher
density clumps.

\item The NH$_3$-derived masses of C1, C2, C3, and C4 are $140 \pm 20$,
$160 \pm 40$, $30 \pm 20$, and $30 \pm 20$~M$_{\sun}$.  The clumps have
masses significantly lower than their virial masses; however, signs
of star formation indicate C1, C2, and C3 are gravitationally bound
and collapsing.  In clumps that are collapsing or contain outflows,
$M_{virial}/M_{core}$ is not a reliable metric for testing gravitational
boundedness.

\item CCS and NH$_3$ are present in each of the three subclouds comprising
G19.30+0.07, and appear to trace different parts of the subclouds.
The NH$_3$ emission closely traces the IR extinction seen in the GLIMPSE
8~\mum\ images, and peaks where clumps are forming and evolving in
the IRDC\@.  Conversely, the CCS tends to trace the boundaries of the
IRDC, where NH$_3$ column densities indicate the IRDC is more diffuse.
Given the high critical density of CCS ($\sim 10^5$) \citep{wolkovitch},
the CCS detections at the IRDC boundaries demonstrate that NH$_3$ does
not give the full picture of the dense gas in star-forming regions.

\item In some places, the CCS appears to have clump-like morphology.
The CCS peaks may show the very earliest stages of clump formation and/or
result from collisions that reset the chemical clock of the gas.

\item Stars are forming in G19.30+0.07, most obviously in clumps C1
and C2.  Younger or lower mass clumps may look like C3, where a H$_2$O
maser but no IR emission is present, or C4, where no signs of star
formation have yet been detected.  S1, where a H$_2$O maser coincides
with a 24~\mum\ source but no NH$_3$ clump is detected, may be a more
evolved protostar, or associated with NH$_3$ at a velocity outside our
observed bandwidth.  We did not detect 8.4~GHz emission from the IRDC,
although the 8.4 GHz emission from the early-B star Orion Source I would
be below our detection limits if it were at a distance of 2.4~kpc.

\end{enumerate}

\section{Acknowledgments}

KD thanks the NRAO pre-doctoral research program for support and
funding.  EC and KD were also supported during this work by the
National Science Foundation under grants AST-0303689 and AST-0808119.
During this investigation RI was supported by a Spitzer Fellowship to
the University of Virginia and by JPL/Spitzer grants RSA 1276990 and
RSA 127546.  The authors thank the anonymous referee for his or her
insightful comments and careful reading of the text.

\begin{deluxetable}{lllll}
\tiny
\rotate
\tablewidth{52pc}
\tablecaption{VLA Observation Details \label{obssum}}
\tablecolumns{5}
\tablehead{
\colhead{Parameter} & \colhead{NH$_3$ (1,1)} & \colhead{NH$_3$ (2,2)}
 & \colhead{CCS ($2_1-1_0$)}  &\colhead{8.4 GHz}\\}
\startdata
Observation Date & 2005 Dec 10, 11 & 2005 Dec 10, 11 & 2007 Feb 19 &
 2006 Jan 30 \\
NRAO Proposal ID & AD516 & AD516 & AD556 & AD524 \\
Rest Frequency (GHz)& 23.6944955 & 23.7226333 & 22.3440330  & 8.4351 \\
VLA Configuration & D & D & D & A \\
Total Bandwidth (MHz) & 3.125 & 3.125 & 3.125 & 50 \\
Number of channels & 128 & 128 & 128 & 1 \\
Spectral Resolution (kHz/\kms) & 24.4/0.3 & 24.4/0.3 & 24.4/0.3 &
 \nodata \\
Primary Beam FWHM & 1.9\arcmin & 1.9\arcmin & 1.9\arcmin & 5.3\arcmin \\
Synthesized Beam & 4\farcs5$\times$3\farcs4& 4\farcs6$\times$3\farcs4 &
 7\farcs7$\times$5\farcs7 & 0\farcs31$\times$0\farcs19 \\
RMS Noise (\mjb\ channel$^{-1}$) & 4.5 & 4.5 & 3 & 0.003 \\
Beam Position Angle & 4.8$^{\circ}$  & $-$0.6$^{\circ}$  & 6.7$^{\circ}$
 & 20.1$^{\circ}$  \\
Conversion Factor (K/Jy) & 144 & 138 & 55.7 & 2.9 $\times 10^5$ \\
Smoothed Synthesized Beam\tablenotemark{*} & 6\farcs0$\times$6\farcs0&
 6\farcs0$\times$6\farcs0 & \nodata & \nodata \\
Smoothed Beam Position Angle\tablenotemark{*} & 0$^{\circ}$ & 0$^{\circ}$
 & \nodata & \nodata \\
RMS Noise (Smoothed Data, \mjb\ channel$^{-1}$) & 6 & 6 & \nodata &
 \nodata \\
Smoothed Conversion Factor (K/Jy) & 60.5 & 60.3 & \nodata & \nodata \\
Largest Detectable Angular Scale & 94\arcsec & 94\arcsec & 100\arcsec &
 16\arcsec \\
Phase and Amplitude Calibrator & J1832$-$105 & J1832$-$105 & J1832$-$105
 & J1832$-$105\\
Flux Density Calibrator & 3C48 & 3C48 & 3C286 & 3C48 \\
Bandpass Calibrator & 3C273  & 3C273 & 3C345 & \nodata \\
\enddata
\tablenotetext{*}{All NH$_3$ data used in line-fitting and
calculations have been spatially smoothed to 6$''$ resolution to
increase the signal to noise ratio.  The only un-smoothed NH$_3$
data are presented in Fig.~\ref{3col19} and in the high-resolution
portions of Fig.~\ref{temperature}a.  It should be further noted that
in Fig.~\ref{ccs}, the NH$_3$ data have been smoothed to match the CCS
beam, 7\farcs7$\times$5\farcs7.}
\end{deluxetable}

\begin{deluxetable}{lcccc}
\small
\tablewidth{36pc}
\tablecaption{ NH$_3$ Clump Properties\label{clumps}}
\tablecolumns{5}
\tablehead{
\colhead{} & \colhead{C1 } & \colhead{C2} & \colhead{C3} & \colhead{C4}
\\}
\startdata
$M_{\rm clump}$ (H$_2$\tablenotemark{*}) (M$_{\sun}$) & $140 \pm 20$ &
 $160 \pm 40$ & $30 \pm 20$ & $30 \pm 20$ \\
$M_{\rm clump}$ (dust\tablenotemark{**}) (M$_{\sun}$) & 113 & 114 &
 \nodata & \nodata \\
$R_{\rm clump}$ (pc\tablenotemark{***}) & $0.11 \pm 0.002$ & $0.12 \pm
 0.003$ & $0.07 \pm 0.006$ & $0.09 \pm 0.005$ \\
$\Delta V$ (\kms) & $2.8 \pm 0.1$ & $1.8 \pm 0.1$ & $1.6 \pm 0.1$ &
 $2.1 \pm 0.2$ \\
$M_{\rm virial}$ (M$_{\sun}$) & $540 \pm 50$ & $260 \pm 40$ & $110 \pm
 20$ & $240 \pm 40$ \\
$M_{\rm virial}$/$M_{\rm clump}$ & $4 \pm 1$ & $2 \pm 1$ & $4 \pm 2$ &
 $8 \pm 5$ \\
\enddata
\tablenotetext{*}{Derived from NH$_3$ as discussed in \S4.4.}
\tablenotetext{**}{Dust results are quoted from \citet{rathborne06}.}
\tablenotetext{***}{$1\arcsec \sim 0.012$~pc assuming a distance of
2.4~kpc.}
\end{deluxetable}

\clearpage

\input{epsf}

\end{document}